\input{epsf}
\documentclass[10.9pt]{article}
\def\be{\begin{equation}}
\def\eea{\end{eqnarray}}
\def\bea{\begin{eqnarray}}
\def\ee{\end{equation}}

\author{M.Mohammadi$^{1}$ \footnote{majid471702@yahoo.com} , M.H.Naderi$^{2}$ \footnote{mhnaderi2001@yahoo.com} and M.Soltanolkotabi$^{2}$ \footnote{soltan@sci.ui.ac.ir}
\\$^{1}$ {\small Department of Physics, Islamic Azad University - Shahreza Branch, Shahreza, Isfahan, Iran}
\\$^{2}$ {\small Quantum Optics Group, Department of Physics, University of Isfahan, Isfahan, Iran}}
\title{Effects of a classical homogeneous gravitational field on the cavity-field entropy and generation of the
 Schr\"{o}dinger-cat states in the Jaynes-Cummings
           model  }
 \begin{document}
\maketitle
\begin{abstract}
\noindent  In this paper, we examine the effects of the
gravitational field on the dynamical evolution of the cavity-field
entropy and the creation of the Schr\"{o}dinger-cat state in the
Jaynes-Cummings model. We consider a moving two-level atom
interacting with a single mode quantized cavity-field in the
presence of a classical homogeneous gravitational field. Based on
an
 su(2) algebra, as the dynamical symmetry group of the model, we derive the reduced density operator of the cavity-field
 which includes the effects of the atomic motion and the gravitational field. Also, we obtain the exact solution and the approximate
 solution for the system-state vector, and examine the atomic dynamics. By considering the temporal evolution of the
 cavity-field entropy as well as the dynamics of the Q-function of the cavity-field we study the effects of the gravitational
 field on the generation of the Schr\"{o}dinger-cat states of the cavity-field by using the Q-function,
  field entropy and approximate solution for the system-state vector. The results show that the
  gravitational field destroys the generation of the Schr\"{o}dinger-cat state of the cavity-field.
\end{abstract}
\noindent PACS numbers: $42.50.c$t$, 42.50.p$q$, 42.50.D$v$ $\\
{\bf Keyword}: Jaynes-Cummings model, atomic motion,
gravitational field, field entropy, Schr\"{o}dinger-cat states\\
\section{Introduction}
The generation of various quantum states occupies a central
position in quantum optics.
 Of special interest are superpositions of coherent states, known as Schr\"{o}dinger-cat states [1].
  Though formed by states having closest classical analogs, such superposition states may exhibit various
  nonclassical properties such as squeezing and sub-Poissonian statistics [2] due to the quantum interference
  between the coherent components. Up to now, several theoretical schemes have been proposed to generate
  various families of such states. Some of those schemes have been addressed to the generation
    of the Schr\"{o}dinger-cat states of the radiation field, for example, in a dispersive medium [3],
   in a Mach-Zehnder interferometer [4], in a Kerr cell [5], in a nanomechanical resonator [6]
    and in a microwave cavity [7]. Theoretical proposals for generating atomic cat states [8]
    and cat states of the motional degree of atoms (ions) [9] have been also considered. Experimentally, the
    Schr\"{o}dinger-cat states of the radiation field have been realized in the context of cavity quantum electrodynamics
    [10] by dispersive coupling between a circular Rydberg atom and the cavity field. Recently, a free-propagating light
    pulse has been also prepared in a Schr\"{o}dinger-cat state [11].\\
    \hspace*{00.5 cm}  In the past two decades, considerable attention
    has been devoted to the dynamical properties of the field entropy [12] and the description of the
    Schr\"{o}dinger-cat states in the Jaynes-Cummings model (JCM) [13]. For the JCM, Phoenix and Knight [14] have
    shown that, at the atomic inversion half-revival time, due to the quantum interaction between the atom and the
    field, the field (atomic) entropy reaches its first minimum value, and the Q-function of the
     field mode bifurcates two blobs having the same amplitude but opposite phase. The asymptotically
     pure field state at this time is approximately a Schr\"{o}dinger-cat state. Buzek and Haldky
     [15] have studied the Schr\"{o}dinger-cat state of the field in the two-photon JCM. They found that
     at quarter of the revival time, the field (atomic) entropy tends to zero, and the Q-function
     also splits into two pieces with the same amplitude but which are out of phase by $\frac{\pi}{2}$. At this time,
     the field is produced in the Schr\"{o}dinger-cat state. However, all these theoretical study
     results are obtained only under the condition that the atomic motion is neglected and the influence
     of the gravitational field is not taken into account.\\
\hspace*{00.5 cm} In the standard JCM, the interaction between a
constant electric field and a stationary (motionless) two-level
atom is considered. With the development in the technologies of
laser cooling and atom trapping the interaction between a moving
atom and the field has attracted much attention [16]. Schlicher
and Joshi [17] have investigated the influences of the atomic
motion and the field-mode structure on the atomic dynamics. They
have shown that the atomic motion and the field-mode structure
give rise to nonlinear transient effects in the atomic population
which are similar to self-induced transparency and adiabatic
effects. Furthermore, the influence of the atomic motion on the
field squeezing in the two-photon Jaynes-Cummings model has been
analyzed [18]. It has been found that the atomic motion does not
destroy the squeezing but decreases the squeezing after a long
time. On the other hand, experimentally, atomic beams with very
low velocities are generated in laser cooling and atomic
interferometry [19]. It is obvious that for atoms moving with a
velocity of a few millimeters or centimeters per second for a time
period of several milliseconds or more, the influence of Earth's
acceleration becomes important and cannot be neglected [20]. For
this reason, it is of interest to study the temporal evolution of
a moving atom simultaneously exposed to the gravitational field
and a single-mode cavity-field. Since any quantum optical
experiment in the laboratory is actually made in a non-internal
frame it is important to estimate the influence of Earth's
acceleration on the outcome of the experiment. By referring to the
equivalence principle, one can get a clear picture of what is
going to happen in the interacting atom-field system exposed to a
classical homogeneous gravitational field [21]. A semi-classical
description of a two-level atom interacting with a running laser
wave in a gravitational field has been studied [22]. However, the
semi-classical treatment does not permit us to study the pure
quantum effects occurring in the course of atom-field interaction.
Recently, within a quantum treatment of the internal and external
dynamics of the atom, we have presented [23] a theoretical scheme
based on an su(2) dynamical algebraic structure to investigate the
influence of a classical homogeneous gravitational field on the
quantum non-demolition measurement of atomic momentum in the
dispersive JCM. Also, we have investigated the effects of the
gravitational field on quantum statistical properties of the
lossless [21] as well as the phase-damped JCMs [24]. We have found
that the gravitational field seriously suppresses non-classical
properties
of both the cavity-field and the moving atom.\\
\hspace*{00.5 cm} The purpose of the present contribution is to
examine the effects of the gravitational field on the dynamical
evolution of the cavity-field entropy and the creation of the
Schr\"{o}dinger-cat state in the JCM. We consider a moving
two-level atom interacting with a single mode quantized
cavity-field in the presence of a classical homogeneous
gravitational field. The atom undergoes one-photon transition
between the nondegenerate states $|g\rangle$ (ground state) and
$|e\rangle$ (excited state). The plan of this paper is as follows.
In section 2, we introduce the physical model and based on an
 su(2) algebra, as the dynamical symmetry group of the model, we obtain an effective Hamiltonian describing
 the atom-field interaction in the presence of gravity. In section 3, we derive the reduced density operator of the cavity-field
 which includes the effects of the atomic motion and the gravitational field. We obtain the exact solution and the approximate
 solution for the system-state vector, and examine the atomic dynamics. In section 4, we consider the temporal evolution of the
 cavity-field entropy as well as the dynamics of the Q-function of the cavity-field and examine the effects of the gravitational
 field on the generation of the Schr\"{o}dinger-cat states of the cavity-field by using the Q-function,
  field entropy and approximate solution for the system-state vector. Finally, we summarize our conclusions in section 5.
\section{The Jaynes-Cummings model in the presence of a gravitational
 field   }
The system considered here consists of a moving two-level atom
interacting with a single-mode quantized cavity-field via the
one-photon transition processes in the presence of gravity. The
total Hamiltonian for the atom-field system in the presence of a
classical homogeneous gravitational field and in the rotating wave
approximation with the atomic motion along the position vector
$\hat{\vec{x}}$ is given by
\begin{eqnarray}
\hat{H}=&&\frac{\hat{p}^{2}}{2M}-M\vec{g}.\hat{\vec{x}}+\hbar\omega_{c}(\hat{a}^{\dag}\hat{a}+\frac{1}{2})+\frac{1}{2}\hbar\omega_{eg}\hat{\sigma}_{z}+\nonumber\\
&&\hbar\lambda[\exp(-i\vec{q}.\hat{\vec{x}})\hat{a}^{\dag}\hat{\sigma}_{-}+\exp(i\vec{q}.\hat{\vec{x}})\hat{\sigma}_{+}\hat{a}],
\end{eqnarray}
where $\hat{a}$ and $\hat{a}^{\dag}$ denote, respectively, the
annihilation and creation operators of the field-mode of frequency
$\omega_{c}$, $\vec{q}$ is the wave vector of the running wave and
$\hat{\sigma}_{\pm}$ denote the raising and lowering operators of
the two-level atom with electronic levels $|e\rangle, |g\rangle $
and Bohr transition frequency $\omega_{eg}$. The atom-field
coupling is given by the parameter $\lambda$ and
 $\hat{\vec{p}}$, $\hat{\vec{x}}$
denote, respectively, the momentum and position operators of the
atomic center of mass motion and $g$ is Earth's gravitational
acceleration. It has been shown [21,23] that based on an su(2)
algebraic structure, as the dynamical symmetry group of the model
and in the interaction picture, the Hamiltonian (1) can be
transformed to the following effective Hamiltonian
\begin{equation}
\hat{\tilde{H}}_{eff}=\hbar \lambda(\sqrt{\hat{K}}\hat{S}_{-}
\exp(-it\hat{\triangle}_{1}(\hat{\vec{p}},\vec{g},t))+\sqrt{\hat{K}}\hat{S}_{+}\exp(it\hat{\triangle}_{1}(\hat{\vec{p}},\vec{g},t))),
\end{equation}
where the operators
\begin{equation}
\hat{S_{0}}=\frac{1}{2}(|e \rangle \langle e|-|g \rangle \langle
g|) , \hat{S_{+}}=\hat{a}|e\rangle \langle
g|\frac{1}{\sqrt{\hat{K}}},
\hat{S_{-}}=\frac{1}{\sqrt{\hat{K}}}|g\rangle \langle
e|\hat{a}^{\dag},
\end{equation}
with the following commutation relations
\begin{equation}
[\hat{S_{0}},\hat{S_{\pm}}]=\pm
\hat{S_{\pm}},[\hat{S_{-}},\hat{S_{+}}]=-2\hat{S_{0}},
\end{equation}
 are the generators of the su(2) algebra, the operator $\hat{K}=\hat{a}^{\dagger}\hat{a}+|e\langle\rangle e|$
 is a constant of motion which represents the total number of excitations of the atom-radiation system and the operator
\begin{equation}
\hat{\triangle}_{1} (\hat{\vec{p}},\vec{g},t)=\frac{1
}{2}(\omega_{c}-(\omega_{eg}+\frac{\vec{q}.\hat{\vec{p}}}{M}+\vec{q}.\vec{g}t+\frac{3\hbar
q^{2}}{2M})),
\end{equation}
is the Doppler shift detuning of the atom-field interaction at
time $t$ [21,23] which depends on both the atomic momentum
and the gravitational field.\\
\section{Dynamical Evolution of the system }
In this section, by using the effective Hamiltonian (2), we
investigate the dynamical evolution of the system under
consideration. At any time $t>0$, the reduced density operator of
the cavity-field is given by
\begin{equation}
\hat{\rho} _{f}(t)=Tr_{a}[\hat{\rho} _{a-f}(t)]=\int d^{3}p
\langle\vec{p} |\otimes(\sum_{i=e,g}\langle i| \hat{\rho}
_{a-f}(t)|i\rangle )\otimes|\vec{p}\rangle,
\end{equation}
where the density operator of the atom-field system is
\begin{equation}
\hat{\rho} _{a-f}(t)=\hat{u}^{\dagger} (t)\hat{\rho} _{a-f}(0)
\hat{u}(t).
\end{equation}
Here, $\hat{\rho}_{a-f}(0)$ is the initial density operator for
the system and $\hat{u}(t)$ is the corresponding time evolution
operator for Hamiltonian (2),
\begin{equation}
\hat{u}(t)=\exp(\frac{-i}{\hbar} \int_{0}^{t}\hat{H}_{eff}(t')dt'
).
\end{equation}
\hspace*{00.5 cm}We assume that initially the radiation field is
prepared in a coherent state, the atom is in the excited state
$|e\rangle$, and the state vector for the center-of-mass degree of
freedom is $|\psi_{c.m}(0)\rangle=\int d^{3}p
\phi(\vec{p})|\vec{p}\rangle$. Therefore, the initial density
operator of the atom-radiation system reads as
\begin{equation}
\hat{\rho}_{a-f}(0)=\hat{\rho}_{f}(0)\otimes
\hat{\rho}_{a}(0)\otimes\hat{\rho}_{c.m}(0),
\end{equation}
where
\begin{equation}
\hat{\rho}_{f}(0)=\sum_{n=0}^{\infty}
\sum_{m=0}^{\infty}w_{n}(0)w_{m}(0)|n\rangle \langle m |,
\end{equation}
\begin{equation}
\hat{\rho}_{c.m}(0)=\int d^{3}p \int
d^{3}p'\phi^{*}(\vec{p'})\phi(\vec{p})|\vec{p}\rangle
\langle\vec{p'} |,
\end{equation}
with
$w_{n}(0)=\frac{\exp(-\frac{|\alpha|^{2}}{2})\alpha^{n}}{\sqrt{n!}}$
and $\phi(\vec{p})=\frac{1}{\sqrt{2\pi
\sigma_{0}}}\exp(\frac{-p^{2}}{\sigma_{0}^{2}})$.\ After some
straightforward calculation, we obtain
\begin{equation}
\hat{\rho}_{f}(t)=|C(t)\rangle\langle C(t)|+|D(t)\rangle\langle
D(t)|,
\end{equation}
where
\begin{equation}
|C(t)\rangle=\int d^{3}p \phi(\vec{p})\sum_{n=0}^{\infty}
w_{n}(0)\sqrt{a_{n}(\vec{p},\vec{g},t)}\exp(\frac{i}{2}E_{+}(\vec{p},\vec{g},t
)\sqrt{n+1})|n\rangle \otimes|\vec{p}\rangle,
\end{equation}
\begin{equation}
|D(t)\rangle=\int d^{3}p \phi(\vec{p})\sum_{n=0}^{\infty}
w_{n-1}(0)\sqrt{b_{n}(\vec{p},\vec{g},t)}\exp(\frac{i}{2}E_{+}(\vec{p},\vec{g},t
)\sqrt{n})|n\rangle \otimes|\vec{p}\rangle,
\end{equation}
with
\begin{equation}
a_{n}(\vec{p},\vec{g},t)=(1-i(n+1)E_{+}(\vec{p},\vec{g},t
)E_{-}^{2} (\vec{p},\vec{g},t )),
\end{equation}
\begin{equation}
b_{n}(\vec{p},\vec{g},t)=i(n+1)E_{+}(\vec{p},\vec{g},t )E_{-}^{2}
(\vec{p},\vec{g},t ),
\end{equation}
and
\begin{eqnarray}
E_{+}(\vec{p},\vec{g},t )=&&(\frac{1}{2}-\frac{i}{2})
\frac{\sqrt{\pi}}{\sqrt{\vec{q}.\vec{g} }
}\exp(-i\frac{\triangle_{0}^{2}(\vec{p})}{2 \vec{q}.\vec{g}} ) (-
Erf[i
(-1)^{\frac{3}{4}}\frac{\triangle_{0}(\vec{p})}{\sqrt{2\vec{q}.\vec{g}
} } ] \\ \nonumber + &&
Erf[(-1)^{\frac{3}{4}}(\frac{\triangle_{0}(\vec{p})}{\sqrt{2\vec{q}.\vec{g}
} }-\sqrt{\frac{\vec{q}.\vec{g} }{2}}t) ] ),
\end{eqnarray}
\begin{eqnarray}
E_{-}(\vec{p},\vec{g},t )=&&(\frac{1}{2}+\frac{i}{2})
\frac{\sqrt{\pi}}{\sqrt{\vec{q}.\vec{g} }
}\exp(i\frac{\triangle_{0}^{2}(\vec{p})}{2 \vec{q}.\vec{g}} ) (-
Erf[i
(-1)^{\frac{3}{4}}\frac{\triangle_{0}(\vec{p})}{\sqrt{2\vec{q}.\vec{g}
} } ] \\ \nonumber + &&
Erf[i(-1)^{\frac{3}{4}}(\frac{\triangle_{0}(\vec{p})}{\sqrt{2\vec{q}.\vec{g}
} }-\sqrt{\frac{\vec{q}.\vec{g} }{2}}t) ] ),
\end{eqnarray}
in which
\begin{equation}
\triangle_{0}(\vec{p})=\triangle_{0}-\frac{\vec{q}.\vec{p}}{2M},
\end{equation}
with
\begin{equation}
\triangle_{0}= \frac{1}{2}[\omega_{c}-(\omega_{eg}+3\frac{\hbar
q^{2}}{2M})],
\end{equation}
is time-independent and $Erf$ denotes the error function. As is
seen, the reduced density operator of the cavity-field includes
the effects of both the atomic motion and the gravitational field.
The state vector for the system can be written as
\begin{equation}
|\psi(t)\rangle=|C(t)\rangle |e\rangle+|D(t)\rangle |g\rangle.
\end{equation}
\hspace*{00.5 cm} Equation (21) is an exact solution for the
system under consideration. Following the work of Gea-Banacloche
[25], for large initial mean photon number $(|\alpha|^{2}>>1)$,
considering the property of Poissonian distribution, we can make
some approximations [26]: $w_{n}(0) \approx w_{n-1}(0)$, $\sqrt{n}
\approx \sqrt{|\alpha|^{2}} [1+\frac{(n-|\alpha|^{2}
)}{2|\alpha|^{2}} ]$, and derive the following approximate
solution for the JCM in the presence of a gravitational field
\begin{equation}
|\psi(t)\rangle\approx|C'(t)\rangle |e\rangle+|D'(t)\rangle
|g\rangle,
\end{equation}
where
\begin{equation}
|C'(t)\rangle \approx \int d^{3}p \phi(\vec{p})\sum_{n=0}^{\infty}
w_{n}(0)\sqrt{a'_{n}(\vec{p},\vec{g},t)}\exp(\frac{i}{2}E_{+}(\vec{p},\vec{g},t
)\sqrt{n+1})|n\rangle \otimes|\vec{p}\rangle,
\end{equation}
\begin{equation}
|D'(t)\rangle \approx \int d^{3}p \phi(\vec{p})\sum_{n=0}^{\infty}
w_{n}(0)\sqrt{b'_{n}(\vec{p},\vec{g},t)}\exp(\frac{i}{2}E_{+}(\vec{p},\vec{g},t
)\sqrt{n})|n\rangle \otimes|\vec{p}\rangle,
\end{equation}
with
\begin{equation}
\sqrt{a'_{n}(\vec{p},\vec{g},t)} \approx
\sqrt{\eta(\vec{p},\vec{g},t) |\alpha|^{2}
}[1+\frac{n+\xi(\vec{p},\vec{g},t)-|\alpha|^{2} }{2|\alpha|^{2}}],
\end{equation}
\begin{equation}
\sqrt{b'_{n}(\vec{p},\vec{g},t)} \approx
\sqrt{-\eta(\vec{p},\vec{g},t) |\alpha|^{2}
}[1+\frac{n+1-|\alpha|^{2} }{2|\alpha|^{2}}],
\end{equation}
 $\eta(\vec{p},\vec{g},t)=-iE_{+}(\vec{p},\vec{g},t)E_{-}^{2}(\vec{p},\vec{g},t)$ and $\xi(\vec{p},\vec{g},t)=1+\eta^{-1}(\vec{p},\vec{g},t)$. This approximate solution is useful in examining the properties and the
generation of the Schr\"{o}dinger-cat states.\\
\hspace*{00.5 cm} By using Eq. (21) we can evaluate the time
evolution of the atomic inversion which takes the form
\begin{equation}
W(t)=\langle \psi(t)|\hat{\sigma}_{z} |\psi(t)\rangle=\langle
C(t)|C(t) \rangle-\langle D(t)|D(t) \rangle,
\end{equation}
where
\begin{eqnarray}
\langle C(t)|C(t) \rangle=&&\int d^{3}p
|\phi(\vec{p})|^{2}\sum_{n=0}^{\infty} w_{n}(0)w_{n}^{*}
(0)\sqrt{a_{n}(\vec{p},\vec{g},t) a^{*}_{n} (\vec{p},\vec{g},t)}\\
\nonumber \times &&\exp(\frac{i}{2}(E_{+}(\vec{p},\vec{g},t
)\sqrt{n+1}- E_{+}^{*} (\vec{p},\vec{g},t )\sqrt{n+1} )),
\end{eqnarray}
\begin{eqnarray}
\langle D(t)|D(t) \rangle=&&\int d^{3}p
|\phi(\vec{p})|^{2}\sum_{n=0}^{\infty} w_{n-1}(0)w_{n-1}^{*} (0)\\
\nonumber \times &&\sqrt{b_{n}(\vec{p},\vec{g},t) b^{*}_{n}
(\vec{p},\vec{g},t)}\exp(\frac{i}{2}(E_{+}(\vec{p},\vec{g},t
)\sqrt{n}- E_{+}^{*} (\vec{p},\vec{g},t )\sqrt{n} )).
\end{eqnarray}
The numerical results of the atomic inversion $W(t)$ are shown in
Fig.1 for initial mean photon number $|\alpha|^{2}=25$ and three
different values of the parameter $\vec{q}.\vec{g}$. In this
figure and all the subsequent figures we set $q=10^{7}m^{-1}$,
$M=10^{-26}Kg$, $g=9.8\frac{m}{s^{2}}$, $\omega_{rec}=\frac{\hbar
q^{2}}{2M}=.5\times10^{6}\frac{rad}{s}$, $\lambda=1\times
10^{6}\frac{rad}{s}$, $\sigma_{0}=1$ and
$\triangle_{0}=8.5\times10^{7}\frac{rad}{s}$
 [21-22]. Fig.1a displays the case when the gravitational influence is negligible.
  This means very small $\vec{q}.\vec{g}$, i.e., the
momentum transfer from the laser beam to the atom is only slightly
altered by the gravitational acceleration because the latter is
very small or nearly perpendicular to the laser beam. Figs.1b and
1c illustrate the case when we consider the gravitational
influence for $\vec{q}.\vec{g}=0.5\times 10^{7}$ and
$\vec{q}.\vec{g}=1.5\times 10^{7}$, respectively. It can be seen
from Fig.1a that, in the condition of no gravitational influence,
the atomic population inversion shows the collapse-revival
repeatedly, and the amplitude of Rabi oscillation in each revival
period is not the same. Increasing of the gravitational influence
(Fig.1b and 1c) induces the population inversion to oscillate so
drastically that the phenomenon of collapse-revival is not so
clear.
\section{Dynamical Properties of the cavity-field}
In this section, we study the time evolution of the cavity-field
entropy as well as the dynamics of the Q-function of the
cavity-field and examine the influence of the gravitational
field on the generation of the Schr\"{o}dinger-cat state. \\
  \\
  \\
  \\
{\bf  4a. Temporal evolution of the field entropy} \\
\\
Here, we use the field entropy as a measure for the degree of
entanglement between the cavity-field and the
 atom of the system under consideration.The quantum dynamics described by the Hamiltonian (1) leads to an entanglement between
 the cavity-field and the atom, which will be quantified by the field entropy. As shown by Phoenix and Knight [14]
 the von Neumann quantum entropy is a convenient and sensitive measure of the entanglement of two interacting subsystems
\begin{equation}
S=-Tr(\hat{\rho}\ln\hat{\rho}),
\end{equation}
where $\hat{\rho}$ is the density operator for a given quantum
system and we have set the Boltzmann constant $k_{B}=1$. If the
atom is in a pure state, then in a suitable basis the density
operator is diagonal and has a single element, unit. For this
case, $S=0$ and if $\hat{\rho}$ describes a mixed state, then
$S\neq 0$. Araki and Lieb [27] showed that these entropies for a
composite system satisfy the triangle inequality $|S_{a} -S_{f}
|\leq S \leq S_{a} +S_{f} $. Quantum entropies are generally
difficult to compute because they involve the diagonalization of
large (and, in many cases, infinite dimensional) density matrices.
Thus explicit illustration of the triangle inequality is
difficult. Phoenix and Knight gave a nice illustration of the
triangle inequality in the context of the JCM. In our model, the
initial state is prepared in a pure state, so the whole atom-field
system remains in a pure state at any time $t>0$ and its entropy
is always zero. However, due to the entanglement of the atom and
the cavity-field at $t>0$, both the atom and the field are
generally in mixed states, although at certain times the field and
the atomic subsystems are almost in pure states. Since the initial
state is a pure state, the entropy $S_{f}$ or $S_{a}$,which is
referred to as the entanglement of the total system in quantum
information, is used to measure the amount of entanglement between
the two subsystems. When  $S_{f}=S_{a}=0$, the system is
disentangled or separable and  both the field and atomic
subsystems are in pure state. The entropies of the particle and
the field, are defined through the corresponding reduced density
operators by
 \begin{equation}
S_{a(f)}=-Tr_{a(f)}(\hat{\rho}_{a(f)}\ln\hat{\rho}_{a(f)}),
\end{equation}
provided we treat both separately. Since the trace is invariant
under a similarity transformation, we can go to a basis in which
the density matrix of the cavity-field is diagonal and then
express the field entropy $S_{f}$ in terms of the eigenvalues
$\pi_{f}^{\pm}(t)$, for the reduced field density operator.
Phoenix and Knight [14] have developed a general method to
calculate the various field eigenstates in a simple way. To
calculate the eigenvalues of the reduced cavity-field density
operator, we write the eigenequation as
  \begin{equation}
\hat{\rho} _{f}(t)|\psi_{f}(t) \rangle=\pi_{f}^{\pm}|\psi_{f}(t)
\rangle.
\end{equation}
The eigenvalues of $\hat{\rho} _{f}(t)$ are
\begin{equation}
\pi_{f}^{\pm}(t)=\frac{1}{2} \pm \frac{1}{2} \sqrt{(1-4(\langle
C(t)|C(t) \rangle \langle D(t)|D(t) \rangle-|\langle
C(t)|D(t)\rangle|^{2}  )},
\end{equation}
where $\langle C(t)|C(t) \rangle$ and $\langle D(t)|D(t) \rangle$
are given, respectively, by (28) and (29), and
\begin{eqnarray}
\langle C(t)|D(t)\rangle=&&\int d^{3}p
|\phi(\vec{p})|^{2}\sum_{n=0}^{\infty} w_{n}(0)w_{n+1}^{*}
(0)\sqrt{a^{*}_{n}(\vec{p},\vec{g},t) b_{n}(\vec{p},\vec{g},t)}\\
\nonumber \times &&\exp(\frac{i}{2}(E_{+}(\vec{p},\vec{g},t
)\sqrt{n+1}- E_{+}^{*} (\vec{p},\vec{g},t )\sqrt{n+2} )).
\end{eqnarray}
The field entropy $S_{f}(t)$ may be expressed in terms of the
eigenvalues $\pi_{f}^{\pm}$ for the reduced field density operator
as
\begin{equation}
S_{f}(t)=-(\pi_{f}^{+}(t)\ln\pi_{f}^{+}(t)+\pi_{f}^{-}(t)\ln\pi_{f}^{-}(t)
).
\end{equation}
As is seen, the field entropy depends on not only the statistics
of the cavity-field but also
the atomic motion and the gravitational field.\\
 \hspace*{00.5 cm} Now we turn our attention to examine numerically the dynamics of the field
 entropy. The numerical results of the evolution of the field entropy are shown in Fig.2 with
 the same corresponding data used in Fig.1. Fig.2a illustrates the case when the gravitational
 influence is negligible, while the rest figures display the cases when we consider the gravitational
 influence. It is seen from Fig.2a that, when the gravitational influence is negligible, the cavity-field
 entropy evolves periodically in the course of time. This periodic evolution can be attributed
 to the atomic motion. Increasing the parameter $\vec{q}.\vec{g}$ (Figs.2b and 2c) results in not only
 increasing in the amplitude of the field entropy but also occurring fast oscillations in the course
 of time evolution of the field entropy.\\
\\
\\
{\bf  4b. Q-function and Schr\"{o}dinger-cat state } \\
\\
Now, we turn our attention to the dynamics of the
quasi-probability distribution Q-function and the analysis of the
generation of the Schr\"{o}dinger-cat states in the JCM in the
presence of the gravitational field. The field entropy and the
Q-function are very useful tools for analyzing the formation of
the Schr\"{o}dinger-cat state. When the field mode is in a state
with the minimum value of the field entropy and the Q-function is
composed of two equal peaks, we can infer that such a field-mode
state is a
Schr\"{o}dinger-cat state and the state vector of the system can be expressed in a factored form at this time.\\
\hspace*{00.5 cm} The Q-function of the cavity-field is defined in
terms of the diagonal elements of the density operator in the
coherent state basis,
\begin{equation}
Q(\beta,\beta^{*},t)=\frac{1}{\pi}\langle\beta|\hat{\rho}_{f}(t)|\beta\rangle,
\end{equation}
where $\beta=X+iY$. By using the explicit expression for the
reduced density operator of the cavity-field given by Eq.(12) we
obtain
\begin{eqnarray}
Q(\beta,\beta^{*},t)=&&\frac{1}{\pi}[|\langle
\beta|C(t)\rangle|^{2}+|\langle \beta|D(t)\rangle|^{2}]\\
\nonumber=&&  \frac{1}{\pi}\int d^{3}p
|\phi(\vec{p})|^{2}\sum_{n}\sum_{m}w_{n}(0)w_{m}^{*} (0)\\
\nonumber \times &&\exp(\frac{i}{2}(E_{+}(\vec{p},\vec{g},t
)\sqrt{n+1}-E_{+}^{*} (\vec{p},\vec{g},t )\sqrt{m+1}))
\\ \nonumber\times&&(\sqrt{a_{n}(\vec{p},\vec{g},t)a^{*}_{m}
\vec{p},\vec{g},t)}\langle \beta|n \rangle\langle m|\beta\rangle
\\ \nonumber + &&\sqrt{b_{n+1}(\vec{p},\vec{g},t)b^{*}_{m+1}
(\vec{p},\vec{g},t)}\langle \beta|n+1 \rangle\langle
m+1|\beta\rangle ).
\end{eqnarray}
\hspace*{00.5 cm} In Fig.3 we have sketched the three-dimensional
plots of the Q-function of the cavity-field versus $X$ and $Y$
 for three values of the
 parameter $\vec{q}.\vec{g}$. The field in coherent state with mean initial photon number
 $|\alpha|^{2}=25$ and the atom in the excited state is considered. Furthermore, we choose $t=t_{R}/2=7\pi/2\lambda$
 which corresponds to one-half of the revival time of the atomic inversion when the gravitational field
 is not taken into account [see Fig.1a]. Fig.3a, which corresponds to the case $\vec{q}.\vec{g}=0$, shows that at $t=t_{R}/2$, the Q-function splits
 into two blobs with the same amplitude but opposite phase. At this time, the state vector for the system can be written
in a factored form approximately by using (22)
\begin{equation}
|\psi(t=t_{R}/2 )\rangle \approx  |\psi_{a} (t=t_{R}/2 )\otimes
|\psi_{f} (t=t_{R}/2 )\otimes|\psi_{c.m} (t=t_{R}/2 ),
\end{equation}
where
\begin{equation}
|\psi_{a}(t=t_{R}/2 )  )=|e\rangle+i|g\rangle,
\end{equation}
\begin{equation}
|\psi_{f}(t=t_{R}/2 )  )=\sum_{n=0}^{\infty}nw_{n}(0)|n\rangle,
\end{equation}
\begin{equation}
|\psi_{c.m}(t=t_{R}/2 )  )=\int
d^{3}p\phi(\vec{p})g(\vec{p},t=t_{R}/2)|\vec{p}\rangle,
\end{equation}
with
\begin{eqnarray}
g(\vec{p},t)=&&[if_{0}^{1/2}(\vec{p},t)\triangle_{0}^{3/2}(\vec{p})\sqrt{|\alpha|^{2}}(1+(1/|\alpha|^{2}))/2\\
\nonumber-&&
2if_{0}^{-1/2}(\vec{p},t)\triangle_{0}^{3/2}(\vec{p})n\sqrt{|\alpha|^{2}}][\cos(A_{0}(\vec{p},t)+F_{0}(\vec{p},t))\\
\nonumber+&& i\sin(A_{0}(\vec{p},t)+F_{0}(\vec{p},t))],
\end{eqnarray}
and
\begin{eqnarray}
f_{0}(\vec{p},t)=&& (\cos(\triangle_{0}(\vec{p})t)-1)^{3}+i
\sin(\triangle_{0}(\vec{p})t)(\cos(\triangle_{0}(\vec{p})t)-1)\\
\nonumber + &&
(\cos(\triangle_{0}(\vec{p})t)-1)\sin(\triangle_{0}(\vec{p})t)^{2}+i
\sin^{3}(\triangle_{0}(\vec{p})t),
\end{eqnarray}
\begin{eqnarray}
A_{0}(\vec{p},t)=&&
\frac{i|\alpha|^{-1}\triangle_{0}(\vec{p})^{-1}}{2}(1+\frac{1-|\alpha|^{2}}{2|\alpha|^{2}})\\
\nonumber \times &&
(\cos(\triangle_{0}(\vec{p})t)-1-i\sin(\triangle_{0}(\vec{p})t)),
\end{eqnarray}
\begin{equation}
F_{0}(\vec{p},t)= \frac{in|\alpha|\triangle_{0}(\vec{p})^{-1}}{4}
(\cos(\triangle_{0}(\vec{p})t)-1-i\sin(\triangle_{0}(\vec{p})t)).
\end{equation}
Combining this result with the fact that the field entropy at this
time tends to zero (Fig.2a), we conclude that at this time the
cavity-field is in the  Schr\"{o}dinger-cat superposition of
macroscopic states. In Figs.3b and 3c we consider the
gravitational influence for $\vec{q}.\vec{g}=0.5\times 10^{7}$ and
$\vec{q}.\vec{g}=1.5\times 10^{7}$, respectively. By comparing
Fig.3a with Figs.3b and 3c, we find that the gravitational field
plays an important role in the dynamics of the Q-function. In
Figs.3b and 3c, two peaks of the Q-function join together and the
state vector for the system cannot be written in a factored form.
For $\vec{q}.\vec{g}=0.5\times 10^{7}$ we have
\begin{eqnarray}
|\psi(t=t_{R}/2)\rangle\approx &&[(\int d^{3}p
\phi(\vec{p})\sqrt{\eta_{1}(\vec{p})}(\sum_{n=0}^{\infty}
w_{n}(0)ng_{1,n}(\vec{p})|n\rangle)\otimes|\vec{p}\rangle)]|e\rangle
\\ \nonumber + && i[(\int d^{3}p
\phi(\vec{p})\sqrt{\eta_{1}(\vec{p})}(\sum_{n=0}^{\infty}
w_{n}(0)ng_{1,n-1}(\vec{p})|n\rangle)\otimes|\vec{p}\rangle)]|g\rangle
\\ \nonumber + &&[(\int d^{3}p
\phi(\vec{p})(26\sqrt{\eta_{1}(\vec{p})}+\eta_{1}^{-1/2}(\vec{p}))\\
\nonumber \times &&(\sum_{n=0}^{\infty}
w_{n}(0)ng_{1,n}(\vec{p})|n\rangle)\otimes|\vec{p}\rangle)]|e\rangle
\\ \nonumber +&& 26i[(\int d^{3}p
\phi(\vec{p})\sqrt{\eta_{1}(\vec{p})}(\sum_{n=0}^{\infty}
w_{n}(0)g_{1,n-1}(\vec{p})|n\rangle)\otimes|\vec{p}\rangle)]|g\rangle
\\ \nonumber \neq && |\psi_{a}(t={\frac{t_R}{2}})\rangle \otimes
|\psi_{f}(t={\frac{t_R}{2}})\rangle \otimes
|\psi_{c.m}(t={\frac{t_R}{2}})\rangle,
\end{eqnarray}
where
\begin{eqnarray}
\eta_{1}(\vec{p})=&& 0.43\times 10^{-9}(1-i)\exp(1 \times
10^{-7}i\triangle_{0}^{2}(\vec{p}))\\ \nonumber \times &&
(-e_{1}(\vec{p})+e_{2}(\vec{p}) )(-e_{1}(\vec{p})+e_{3}(\vec{p})
)^{2},
\end{eqnarray}
\begin{eqnarray}
g_{1,n}(\vec{p})=&& 0.1\exp[0.5i\sqrt{n}(0.38\times 10^{-3}(1-i) \\
\nonumber \times
&&\exp[-10^{-7}i\triangle_{0}^{2}](-e_{1}(\vec{p})+e_{2}(\vec{p})))],
\end{eqnarray}
with
\begin{equation}
e_{1}(\vec{p})=Erf[-0.3 \times 10^{-3} \sqrt{i}
\triangle_{0}(\vec{p})],
\end{equation}
\begin{equation}
e_{2}(\vec{p})=Erf[(0.3 i \sqrt{i}
\triangle_{0}(\vec{p})-17.6i\sqrt{i})10^{-3}],
\end{equation}
\begin{equation}
e_{3}(\vec{p})=Erf[(-0.3 \sqrt{i}
\triangle_{0}(\vec{p})+17.6\sqrt{i})10^{-3}],
\end{equation}
and for $\vec{q}.\vec{g}=1.5\times10^{7}$ we have
\begin{eqnarray}
|\psi(t=t_{R}/2)\rangle\approx &&[(\int d^{3}p
\phi(\vec{p})\sqrt{\eta_{2}(\vec{p})}(\sum_{n=0}^{\infty}
w_{n}(0)ng_{2,n}(\vec{p})|n\rangle)\otimes|\vec{p}\rangle)]|e\rangle
\\ \nonumber + && i[(\int d^{3}p
\phi(\vec{p})\sqrt{\eta_{2}(\vec{p})}(\sum_{n=0}^{\infty}
w_{n}(0)ng_{2,n-1}(\vec{p})|n\rangle)\otimes|\vec{p}\rangle)]|g\rangle
\\ \nonumber + &&[(\int d^{3}p
\phi(\vec{p})(26\sqrt{\eta_{2}(\vec{p})}+\eta_{2}^{-1/2}(\vec{p}))\\
\nonumber \times &&(\sum_{n=0}^{\infty}
w_{n}(0)ng_{2,n}(\vec{p})|n\rangle)\otimes|\vec{p}\rangle)]|e\rangle
\\ \nonumber +&& 26i[(\int d^{3}p
\phi(\vec{p})\sqrt{\eta_{2}(\vec{p})}(\sum_{n=0}^{\infty}
w_{n}(0)g_{2,n-1}(\vec{p})|n\rangle)\otimes|\vec{p}\rangle)]|g\rangle
\\ \nonumber \neq && |\psi_{a}(t={\frac{t_R}{2}})\rangle \otimes
|\psi_{f}(t={\frac{t_R}{2}})\rangle \otimes
|\psi_{c.m}(t={\frac{t_R}{2}})\rangle,
\end{eqnarray}
where
\begin{eqnarray}
\eta_{2}(\vec{p})=&& 0.2\times 10^{-9}(1-i)\exp(0.33 \times
10^{-7}i\triangle_{0}^{2}(\vec{p}))\\ \nonumber \times &&
(-e'_{1}(\vec{p})+e'_{2}(\vec{p})
)(-e'_{1}(\vec{p})+e'_{3}(\vec{p}) )^{2},
\end{eqnarray}
\begin{eqnarray}
g_{2,n}(\vec{p})=&& 0.1\exp[0.5i\sqrt{n}(0.22\times 10^{-3}(1-i) \\
\nonumber \times
&&\exp[-0.33\times10^{-7}i\triangle_{0}^{2}](-e_{1}(\vec{p})+e_{2}(\vec{p})))],
\end{eqnarray}
with
\begin{equation}
e'_{1}(\vec{p})=Erf[-0.18 \times 10^{-3} \sqrt{i}
\triangle_{0}(\vec{p})],
\end{equation}
\begin{equation}
e'_{2}(\vec{p})=Erf[(0.18 i \sqrt{i}
\triangle_{0}(\vec{p})-29.9i\sqrt{i})10^{-3}],
\end{equation}
\begin{equation}
e'_{3}(\vec{p})=Erf[(-.18 \sqrt{i}
\triangle_{0}(\vec{p})+29.9\sqrt{i})10^{-3}].
\end{equation}
Combining this result with the fact that in the presence of
gravity the field entropy reaches its maximum value at $t=t_{R}/2$
(Figs.2b and 2c) we find that the cavity-field is in a statistical
mixture state. Therefore, we can conclude that, in the JCM, if the
gravitational field influence is taken into account, there are no
Schr\"{o}dinger-cat states in the course of time evolution of the
cavity-field. In other words, the gravitational field destroys the
generation of the Schr\"{o}dinger-cat state.
\\
\section{Summary and conclusions}
In this paper, we considered a moving two-level atom interacting
with a single mode quantized cavity-field in the presence of a
classical homogeneous gravitational field. Based on an
 su(2) algebra, as the dynamical symmetry group of the model, we obtained the exact and the approximate
 solutions for the system-state vector, and examined the influence of the gravitational field on the evolution of the atomic
 inversion, the field entropy and the generation of the Schr\"{o}dinger-cat state.
 The results are summarized as follows: 1) The gravitational field leads to the occurence of fast oscillations
  in the atomic population inversion such that the collapse and revivals
  phenomena can not be identified clearly.
 2) Increasing the parameter
$\vec{q}.\vec{g}$ results in not only
 increasing in the amplitude of the field entropy but also occurring fast oscillations in the course
 of time evolution of the field entropy. 3) The gravitational field destroys the generation of the Schr\"{o}dinger-cat state.
\\
  \\
\\

\vspace{20cm}

{\bf FIGURE CAPTIONS:}

{\bf FIG. 1 } Time evolution of the atomic population inversion
versus the scaled time $\lambda t$. Here we have set
$q=10^{7}m^{-1}$,\\
$M=10^{-26}kg$,$g=9.8\frac{m}{s^{2}}$,$\omega_{rec}=.5\times10^{6}\frac{rad}{s}$,\\$\lambda=1\times
10^{6}\frac{rad}{s}$,
$\triangle_{0}=8.5\times10^{7}\frac{rad}{s}$ ;\\

 {\bf a)}For $\vec{q}.\vec{g}=0$.

{\bf b)}For $\vec{q}.\vec{g}=0.5 \times 10^{7}$.

{\bf c)}For $\vec{q}.\vec{g}=1.5 \times 10^{7}$.\\

{\bf FIG. 2 } Time evolution of the von Neumann entropy $S$ versus
the scaled time $\lambda t$ with the same corresponding data
 used in Fig.1;\\

{\bf $a$)} For $\vec{q}.\vec{g}=0$.

{\bf $b$)} For $\vec{q}.\vec{g}=0.5 \times 10^{7}$.

{\bf $c$)} For $\vec{q}.\vec{g}=1.5 \times 10^{7}$.

{\bf FIG. 3 } The Q-function of the cavity-field versus
$X=Re(\beta)$ and $Y=Im(\beta)$  with the same corresponding data
 used in Fig.1 and $t={\frac{t_{R}}{2}}$;\\

 {\bf a)}For $\vec{q}.\vec{g}=0$.

{\bf b)}For $\vec{q}.\vec{g}=0.5 \times 10^{7}$.

{\bf c)}For $\vec{q}.\vec{g}=1.5 \times 10^{7}$.\\



\end{document}